\documentclass{ws-procs9x6}

\begin{document}

\title{First Results from AMANDA using the TWR System}

\author{ANDREA SILVESTRI$^{\dagger}$ for the IceCube Collaboration\footnote{http://www.icecube.wisc.edu}}

\address{Department of Physics and Astronomy, University of California\\
Irvine, CA 92697, U.S.A.\\
$^{\dagger}$E-mail: silvestr@uci.edu\\
}

\begin{abstract}
The Antarctic Muon And Neutrino Detector Array (AMANDA) has been
taking data since 2000 and its data acquisition system was
upgraded in January 2003 to read out the complete digitized waveforms
from the buried Photomultipliers (PMTs) using Transient Waveform
Recorders (TWR).
This system currently runs in parallel with the standard AMANDA data
acquisition system. 
Once AMANDA is incorporated into the 1 km$^{3}$ detector IceCube, only
the TWR system will be kept.
We report results from a first atmospheric
neutrino analysis on data collected in 2003 with TWR. Good agreement
in event rate and angular distribution verify the performance of the TWR
system. A search of the northern hemisphere for localized event
clusters shows no statistically significant excess, thus a flux limit
is calculated, which is in full agreement with previous results based on
the standard AMANDA data acquisition system.
We also update the status of a search for diffusely distributed
neutrinos with ultra high energy (UHE) using data collected by the TWR
system.
\end{abstract}


\bodymatter

\section*{Introduction}
Neutrinos are the only high energy particles able to propagate undeflected 
and unattenuated from the furthest reaches of the Universe. Extragalactic
UHE $\gamma$-ray astronomy falters for energies greater than a few tens of 
TeV due to interactions with infrared and Cosmic Microwave Background 
photons. The information carried by the neutrino messengers from distant,
unexplored regions of the universe may help to unravel longstanding 
mysteries associated with the origin of the highest energy cosmic rays.  
Several
models{\small{\cite{Stecker92,Protheroe:1996,yos98,Sigl98,Engel:2001hd,Stecker05}}}
predict high energy charged particle
and neutrino emission; the diffuse $\nu$-flux predictions by these models 
are constrained by the observed cosmic ray
fluxes{\small{\cite{Waxman_99}}}.
AMANDA{\small{\cite{silvestri1}}}, the first neutrino telescope
constructed in transparent ice, is deployed between 1500 m and 2000 m
beneath the surface at the geographic South Pole.
It is designed to search for neutrinos
that originate in the most violent phenomena in the observable universe.
Galactic objects like Supernova Remnants (SNR) and extragalactic
objects such as Active Galactic Nuclei (AGN) are expected to be the
powerful engines
accelerating protons and nuclei to the highest energies, which
eventually interact to generate neutrinos.
We have searched for point sources in the northern sky and for a
diffuse flux of UHE (E$_\nu$ $> 10^{15}$ eV) neutrinos of cosmic origin. 
We have performed the analysis using data collected in 2003 with
a new data acquisition system based on
Transient Waveform Recorders ({\tt TWR-DAQ}), which we compared to data
collected by the standard AMANDA readout system ({$\mu$-{\tt DAQ}}).
\begin{figure}[h]
\begin{center}
\includegraphics*[width=0.70\textwidth,angle=0,clip]{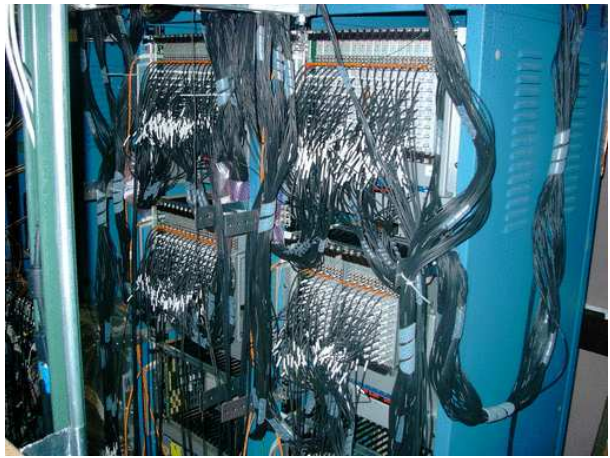}
\caption{\label {fig1} 
The upgraded AMANDA data acquisition electronics with Transient
Waveform Recorders, here displayed in a system of 2 racks with 72 TWR
modules.}
\end{center}
\end{figure}

\section{The TWR System}
The data acquisition electronics of the AMANDA-II detector
(Fig.~\ref{fig1}) was
upgraded in 2003 to read out the complete digitized waveform of the
photomultiplier tubes (PMTs) using Transient Waveform Recorders
(TWR){\small{\cite{icrc2003}}}.
The transition was made to run the
{$\mu$-{\tt DAQ}}{\small{\cite{icrc2001}}} and {\tt TWR-DAQ} in
parallel to verify that the performance of the latter is as good or
better.
To compare the two systems, the data from 2003 has been
analyzed with data sets from both readout systems.
The performance of the {\tt TWR-DAQ} is verified by comparing the
results for the absolute rate of atmospheric neutrinos and the
$cos(\theta)$ distribution with the results
from the standard {$\mu$-{\tt DAQ}}{\small{\cite{silvestri3}}}.
Extending the analysis tools to include {\tt TWR} data required
several new developments:
(1) The {\tt TWR-DAQ} measures the integrated charge
$\cal{Q}$ from the full waveform of the PMT pulses, in contrast, the
{$\mu$-{\tt DAQ}} only measures the maximum amplitude of the PMT pulse
in a 2 $\mu$s window.
(2) Various time offsets were taken into account for the
{\tt TWR}-system, none of which impacted the {$\mu$-{\tt DAQ}}. A time
resolution of few nanoseconds was extracted.
(3) The size of the {\tt TWR-DAQ} data set (20TB for 2003) by far
exceeds that of the {$\mu$-{\tt DAQ}} ($\sim$1TB), requiring an improved
data handling tools.

\section{Analysis}
\label{sec2}
\noindent
The data information of the two systems has been merged
according to GPS time and the fraction of overlapping PMTs
participating in the event in both systems.
For the point source analysis we restricted the capabilities of the
{\tt TWR-DAQ} system to mimic the features of the {$\mu$-{\tt DAQ}}
system as closely as possible. 
The timing and amplitude information extracted from waveforms has been
used as input parameters to perform PMT-pulse cleaning,
TOT (Time-Over-Threshold) and 
crosstalk\footnote{Any phenomenon by which a signal transmitted on one
  channel of a transmission system creates an undesired
  effect in another channel.} cleaning.
\begin{figure}[h]
\begin{center}
\includegraphics*[width=0.49\textwidth,angle=0,clip]{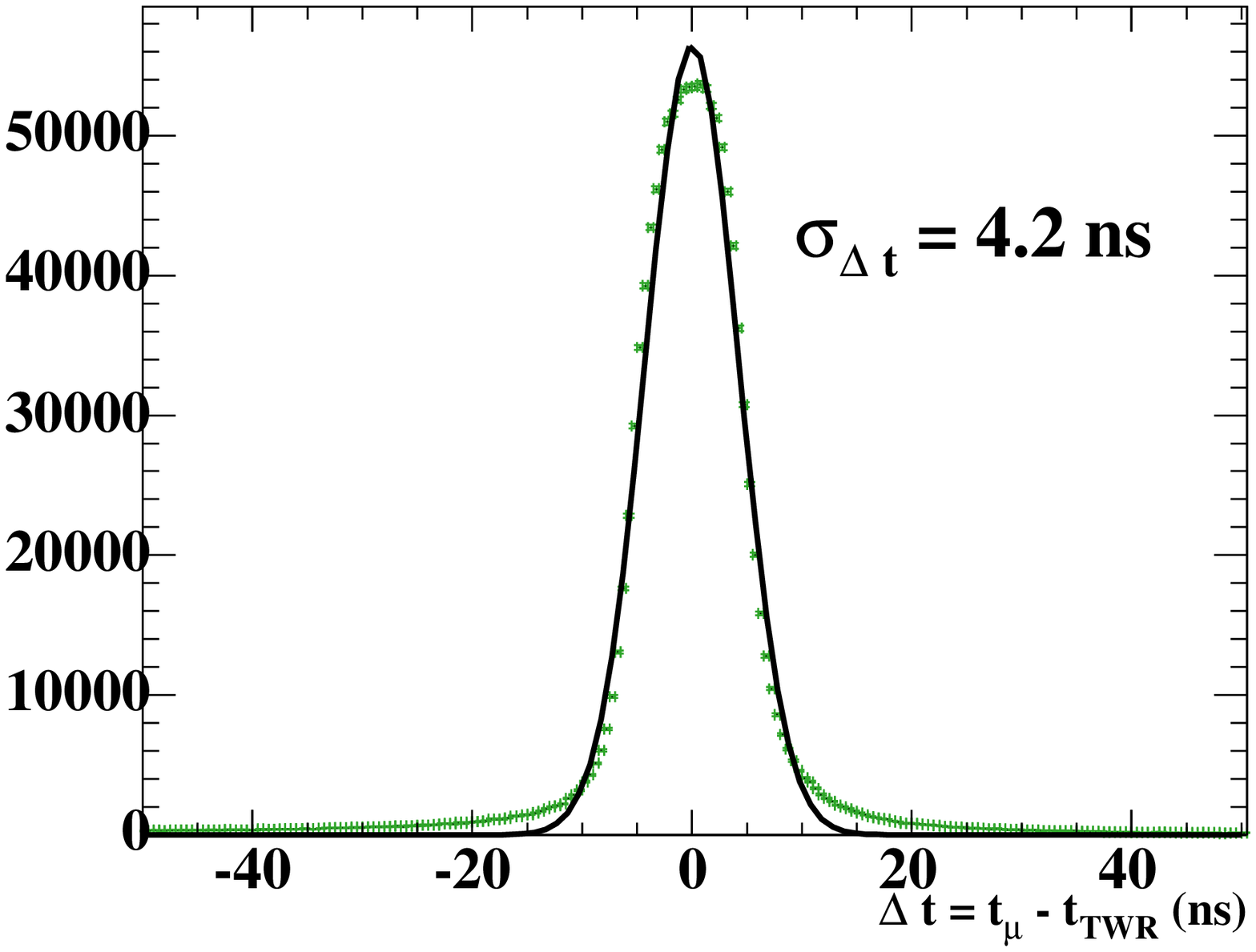}
\includegraphics*[width=0.49\textwidth,angle=0,clip]{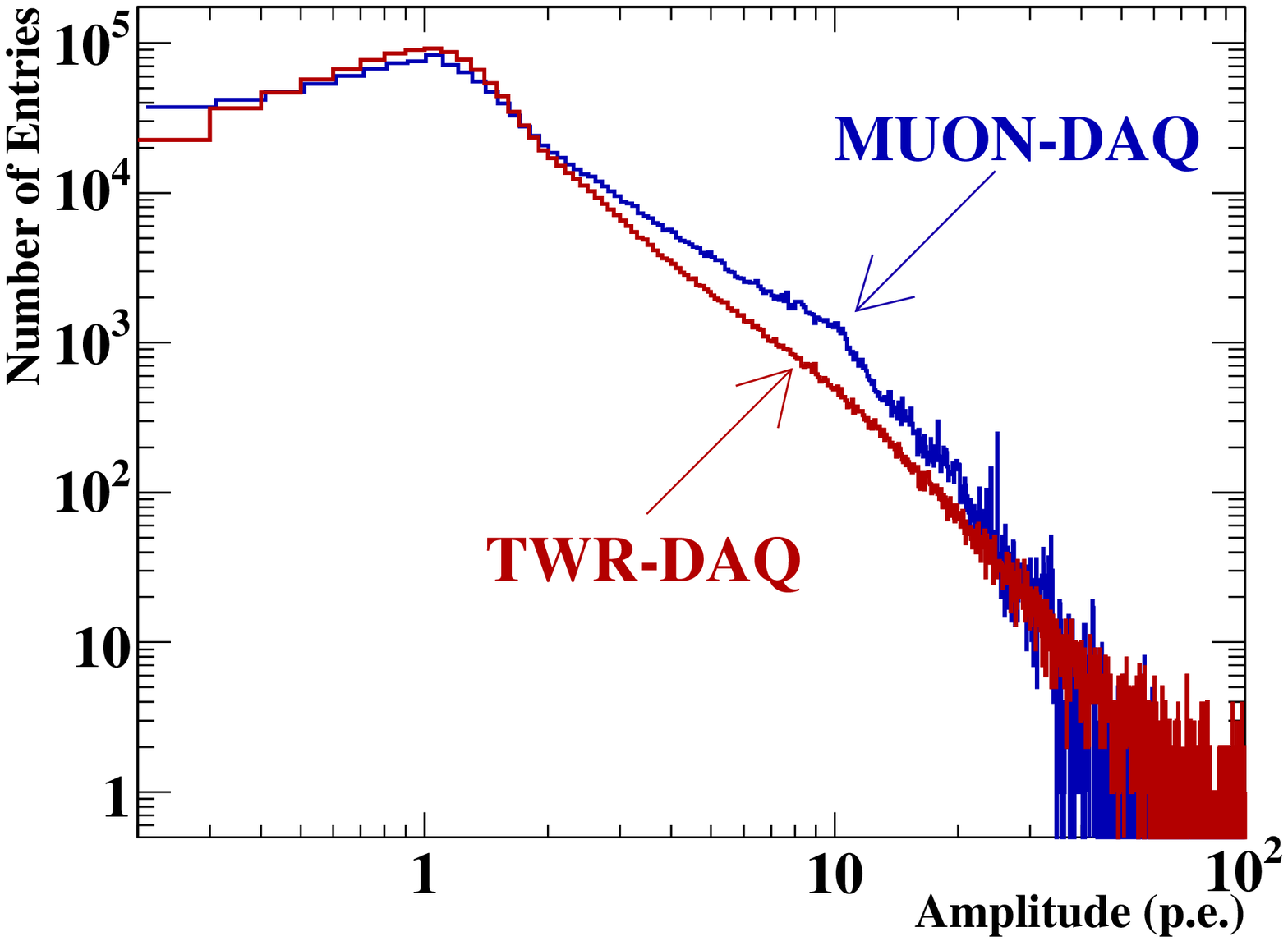}
\caption{\label {fig2} 
(Left) Time difference between the PMT pulses recorded by the two
  acquisition systems $\Delta t = t_{\mu} - t_{TWR}$. (Right)
  Calibrated amplitude normalized to 1 photo-electron (p.e.)
  value for {$\mu$-{\tt DAQ}} and {\tt TWR-DAQ} data. See text for details.}
\end{center}
\end{figure}
A Gaussian fit of the distribution for 
$\Delta t = t_{\mu} - t_{TWR}$ yields $\sigma_{\Delta t} = 4.2$ ns
(Fig. \ref{fig2} (left)), which is dominated by the systematic error
of the time jitter between independent flash ADC clocks of the TWR
system.
The {\tt TWR-DAQ} timing calculations are relative to the
values measured by the {$\mu$-{\tt DAQ}}.
The timing of the {\tt TWR-DAQ} system includes two sources of jitter,
the digitization window of 10 ns and the time fluctuation of the flash
ADC clocks within the same 10 ns interval.
This accounts for the $\sigma_{\Delta t} \sim 4$ ns\footnote{For year
  2005 the phases of the TWR modules were synchronized to avoid the
  time jitter between independent flash ADC clocks.}.
Amplitudes are also calibrated by extracting the number of
photo-electrons (Npe)
detected from the peak ADC of the {$\mu$-{\tt DAQ}} and charge $\cal{Q}$
of the TWR system and normalized to 1-pe amplitude. 
By integrating the charge from pulses in the waveform, the dynamic
range of the {\tt TWR-DAQ} extends to Npe of $\sim 100$, a factor of
three higher than the {$\mu$-{\tt DAQ}}.
Fig. \ref{fig2} (right) shows the reconstructed amplitude of the 
{\tt TWR-DAQ} compared to the {$\mu$-{\tt DAQ}}, which indicates a stable
power law distribution extending up to 100 Npe,
while the {$\mu$-{\tt DAQ}} system shows a ``knee'' around 10 Npe.
The knee is due to the amplitude saturation of
channels read out by optical fibers, which comprise $\sim$40\% of the
AMANDA channels.\footnote{High voltage values were lowered in
January 2005 to increase linear dynamic range of optical channels.}
After cleaning, the muon track is reconstructed from the pulse times
of the PMTs. Details on the reconstruction techniques can be found
in{\small{\cite{amandareco}}}.
\begin{table}
\tbl{Passing rates for increasing level of data selection criteria for
  the {\tt TWR-DAQ} and {$\mu$-{\tt DAQ}} data analysis.}
{\begin{tabular}{@{}ccc@{}}\toprule
~Selection~&~{\tt TWR-DAQ}~ & ~{$\mu$-{\tt DAQ}}~\\
\colrule
Level-0 (Raw)    \hphantom{0} &  $1.86\times10^{9}$ \hphantom{0} & $1.86\times10^{9}$\\
Level-1       \hphantom{0} & $1.25\times10^{8}$ \hphantom{0} &$1.25\times10^{8}$   \\
Level-2       \hphantom{0} & $2.56\times10^{6}$\hphantom{0} & $1.99\times10^{6}$   \\
Level-3 (Final)  \hphantom{0} & 1112\hphantom{0} & 1026\\
\botrule
\end{tabular}
}
\label{tab1}
\end{table}
Table~\ref{tab1} summarizes the passing rates from the raw data level 
to the final sample of atmospheric neutrinos.
The event selection criteria used in this analysis follows the method
described in{\small{\cite{ps2006}}}.
Fig. \ref{fig4} shows that the angular mismatch for the final neutrino sample 
$\sigma_{\Delta \theta}~{\rm is}~0.5^{\circ}$ where $\Delta \theta = \theta_{\mu}
- \theta_{TWR}$.
This value is expected from studies of the precision of the
global minimizer in the reconstruction program, by reconstructing the
same event sample twice with a 32-iteration algorithm. The azimuth
angular mismatch $\sigma_{\Delta \phi}~{\rm is}~0.7^{\circ}$,
which is consistent with a $\sigma$ of $0.5^{\circ}$ for the
$\Delta \phi \times {\rm sin}(\theta)$ distribution to account for the
zenith dependence on the azimuth angle.
\begin{figure}[h]
\begin{center}
\includegraphics*[width=0.49\textwidth,angle=0,clip]{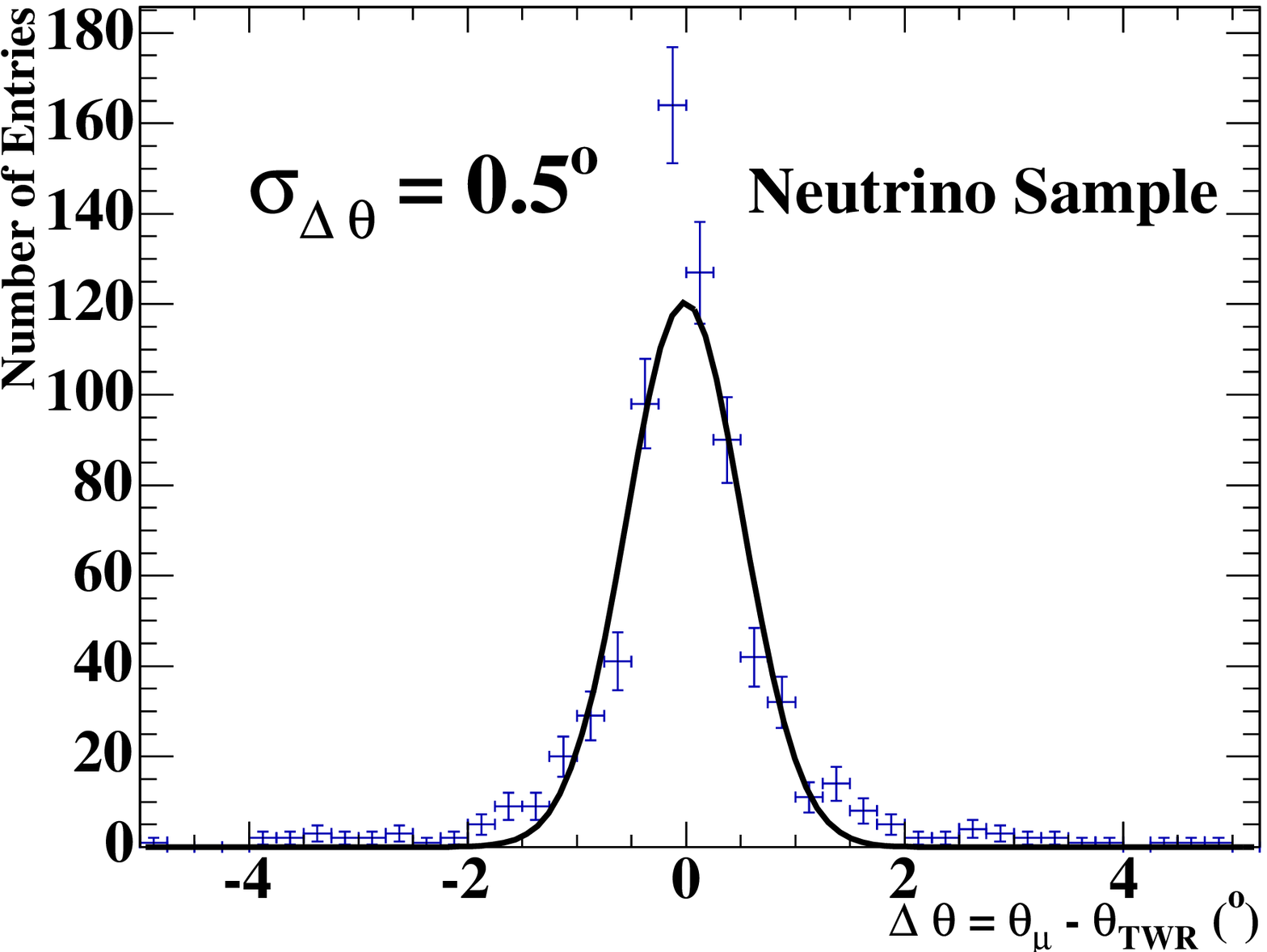}
\includegraphics*[width=0.49\textwidth,angle=0,clip]{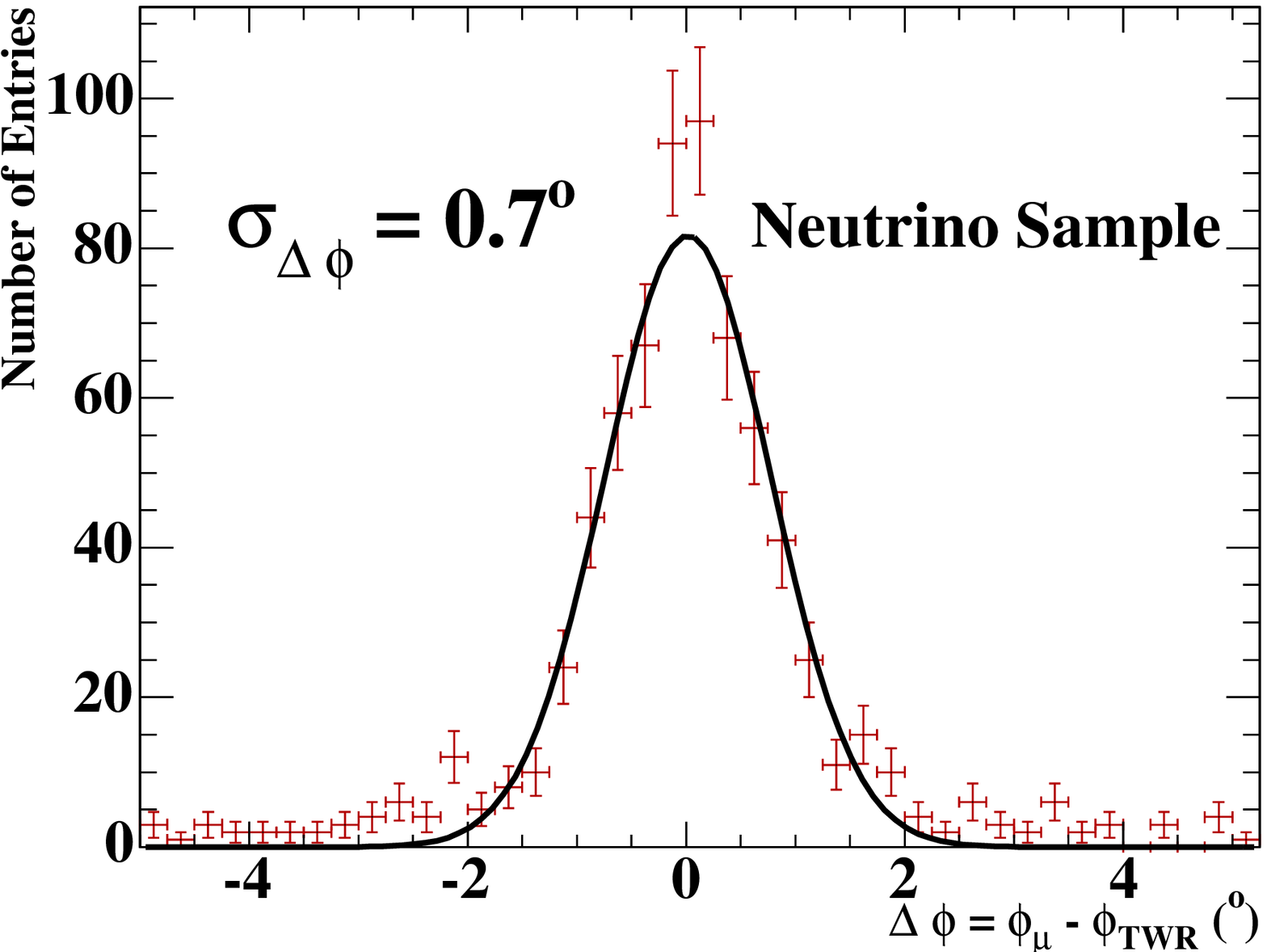}
\caption{\label {fig4} (Left) The zenith $\Delta \theta = \theta_{\mu} -
  \theta_{TWR}$ difference distribution between the {$\mu$-{\tt DAQ}}
  and the {\tt TWR-DAQ} systems, (right) the azimuthal $\Delta \phi = \phi_{\mu}
  - \phi_{TWR}$ difference distribution.}
\end{center}
\end{figure}

\section{Search for Point Sources of Neutrinos}
From the analysis based on {\tt TWR-DAQ} data,
1112 neutrinos are observed compared to 1026 neutrinos from the
{$\mu$-{\tt DAQ}} data analysis.
The small differences in the
event rate are compatible with the small differences in analysis
procedures described in Section \ref{sec2}.
\begin{figure}[h]
\begin{center}
\includegraphics*[width=0.99\textwidth,angle=0,clip]{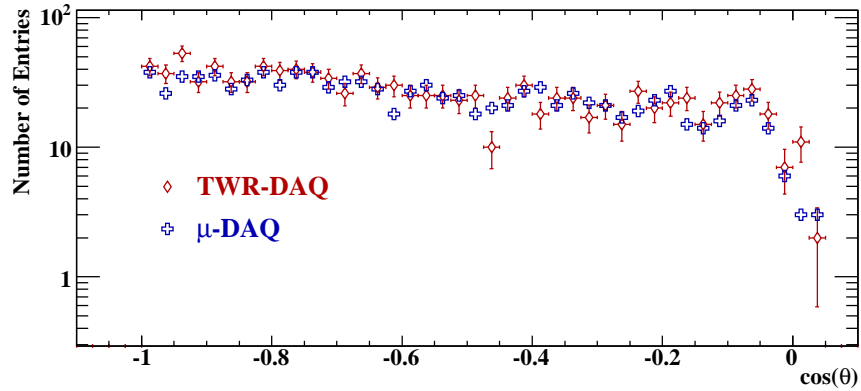}
\caption{\label {fig6} Distribution of cos$(\theta)$ after final
  selection representing the atmospheric neutrino sample observed from
  the {\tt TWR-DAQ} and the {$\mu$-{\tt DAQ}} analyses.}
\end{center}
\end{figure}
\begin{figure}[ht]
\begin{center}
\includegraphics*[width=1.0\textwidth,angle=0,clip]{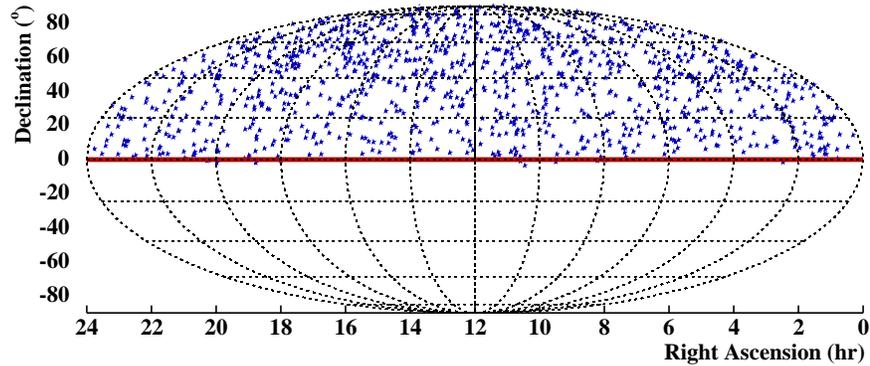}
\caption{\label {fig7} Sky plot displayed in equatorial coordinates,
  Right Ascension (RA) and Declination (DEC), of the final
  sample of 1112 atmospheric neutrinos observed from the {\tt TWR-DAQ}
  analysis.}
\end{center}
\end{figure}
Fig. \ref{fig6} shows the cos$(\theta)$ distribution of the
atmospheric neutrino sample extracted from the {\tt TWR-DAQ} and
{$\mu$-{\tt DAQ}} data analysis.
Satisfactory agreement can be seen for the
cos$(\theta)$ distribution of the atmospheric neutrinos samples obtained
by the different analyses.
Fig. \ref{fig7} displays the sky map plotted in equatorial
coordinates of the 1112 atmospheric neutrino candidates observed from
the {\tt TWR-DAQ} analysis. All events are distributed in
the norther hemisphere with very few events reconstructed at the
horizon. 
In order to distinguish if the observed events follow a
random distribution as expected by atmospheric neutrinos, or are an
indication of localized event cluster as expected by a source of
\begin{figure}[h]
\begin{center}
\includegraphics*[width=1.0\textwidth,angle=0,clip]{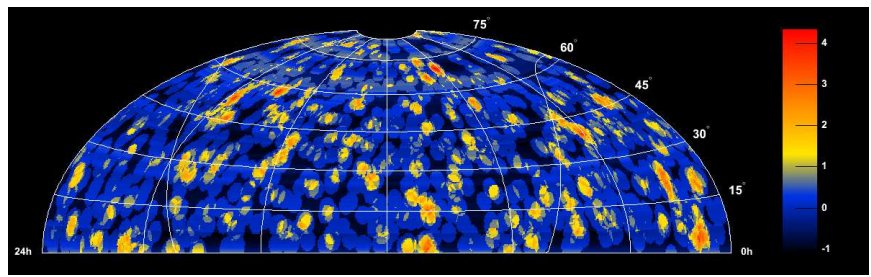}
\caption{\label {fig8}  Sky map plotted in coordinates of
    right ascension and declination of the 1112 atmospheric neutrino
    candidates from the {\tt TWR-DAQ} analysis.
    The scale on the right reflects excess or deficit in
    terms of the positive or negative deviation with respect to the
    mean background events.}
\end{center}
\end{figure}
neutrinos of extraterrestrial origin, an analysis is required to
estimate the statistical significance of the observed events.
A full search of the northern sky was performed to look for any
localized event cluster. The full scan is extended to $85^{\circ}$ in
declination, since the limited statistics in the polar bin prevents an
accurate estimate of the background.
Fig. \ref{fig8} shows the sky map of the calculated
significance for all observed 1112 events in terms of the a positive
or negative deviation with respect to the mean background events. All
observed regions with the highest significance are compatible with the
background hypothesis.
The highest significance observed shows a positive
deviation of 4.3 $\sigma$, which corresponds to a probability of 23\%
for a search performed on events with randomized right ascension.

\begin{figure}[h]
\begin{center}
\includegraphics*[width=0.49\textwidth,angle=0,clip]{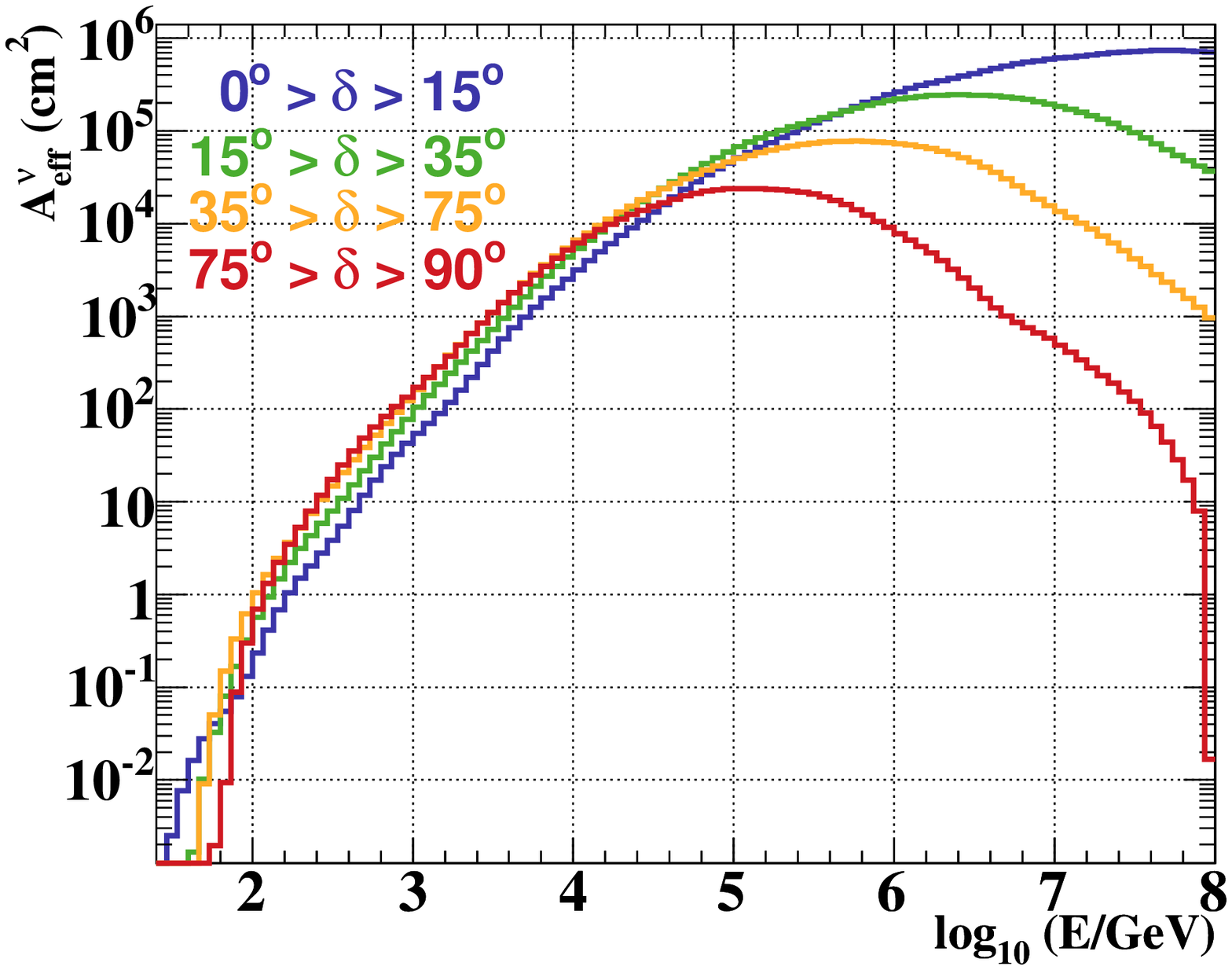}
\includegraphics*[width=0.49\textwidth,angle=0,clip]{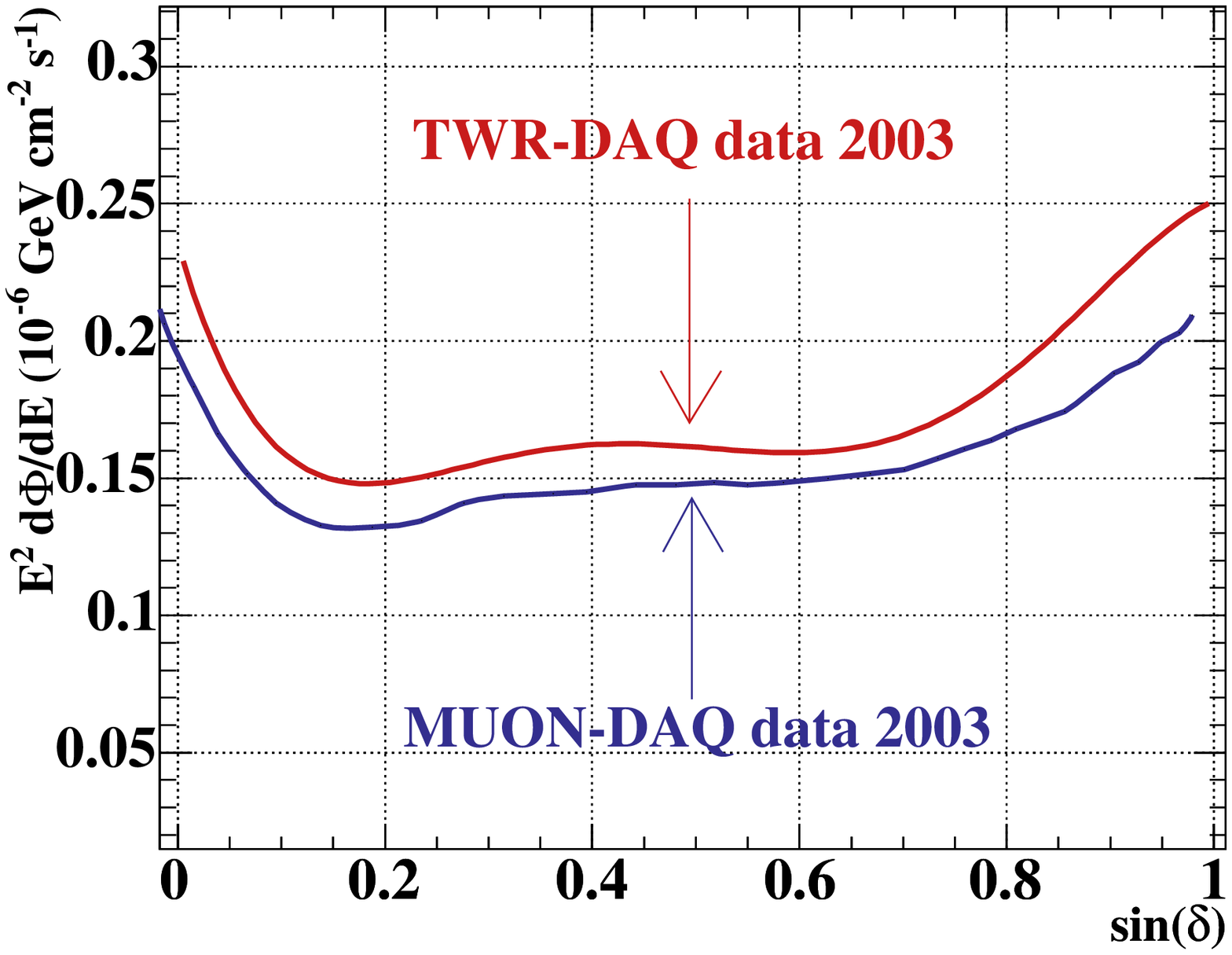}
\caption{\label {fig9}
(Left) Muon neutrino effective area as a function of E$_{\nu_{\mu}}$
  for point source searches displayed for different declinations $\delta$.
(Right) Average upper limit as a function of ${\rm sin}(\delta)$ on $\nu_{\mu}$ for
  a E$^{-2}$ neutrino flux spectrum from the {\tt TWR-DAQ} and the
  {$\mu$-{\tt DAQ}} analyses.}
\end{center}
\end{figure}
\noindent
Numerous studies have been performed to implement the best possible
description of the TWR system in the Monte Carlo simulation.
In particular, a correct description of the waveform
hardware response is now available and proper TDC time windows as well
as amplitude calculations have been now implemented.
Therefore a calculation of the neutrino effective area for point
source search using TWR data is now feasible. Fig.~\ref{fig9} (left)
shows the effective muon neutrino area as a function of
E$_{\nu_{\mu}}$ computed at the final selection of this analysis for
four separated ranges of the declination. The A$^{\nu}_{eff}$ extends
over six order of magnitudes as $\nu$-energy increases, while
A$^{\nu}_{eff}$ decreases as declination increases due to the neutrino
absorption in the Earth.
The average upper limit on $\nu_{\mu}$ as a function of declination is shown
in the right panel of Fig.~\ref{fig9}. The average neutrino flux upper limit
is determined from the ratio of the average Feldman and
Cousins{\small\cite{feldman}} upper limit $<\mu^{90}>$ computed
according to the expected mean background and observed events, and the
number of the expected signal events for a neutrino flux
E$^{2}_{\nu}$d$\Phi_{\nu}/$dE$_\nu =$ GeV s$^{-1} $cm$^{-2}$ from a
point source at declination $\delta$.
Together with the average upper limit extracted from the {\tt TWR-DAQ}
data, the limit from the {$\mu$-{\tt DAQ}} data
of year 2003{\small\cite{ps2006}} is also represented for comparison. These results are
comparable, however the {\tt TWR-DAQ} limit is $\sim$ 10\% worse than
the {$\mu$-{\tt DAQ}} limit, because conservative event selection
criteria have been applied that do not use the full waveform
information in an optimal way.
The preliminary average upper limit from the {\tt TWR-DAQ} analysis on
the muon neutrino flux with
spectrum d$\Phi_{\nu}/$dE$_\nu$ $\propto$ E$^{-2}_{\nu}$ is
E$^{2}_{\nu}$d$\Phi_{\nu}/$dE$_\nu$ $\leq 1.8 \times 10^{-7}$ GeV
s$^{-1} $cm$^{-2}$ at 90\% confidence level, in the energy range
1.26 TeV $<$ E$_{\nu}$ $<$ 1.6 PeV.
This upper limit does not include systematic uncertainties.
AMANDA has submitted a publication on a 5-year search for point sources
(2000-2004) based on the {$\mu$-{\tt DAQ}} data{\small\cite{ps2006}},
which improves the limit by $\sim$67\% compared to the results from
this analysis.
\begin{figure}[h]
\begin{center}
\includegraphics*[width=0.49\textwidth,angle=0,clip]{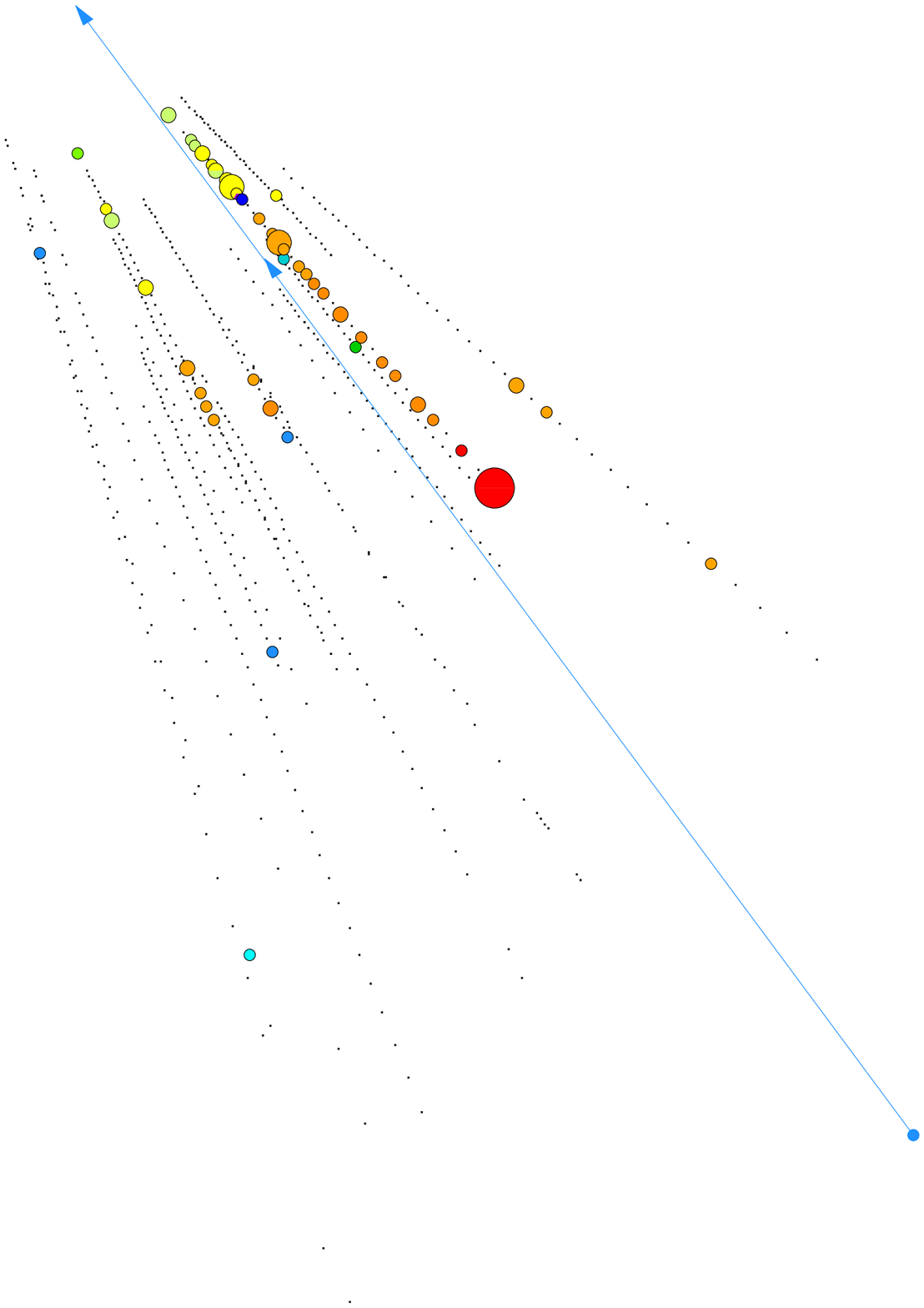}
\includegraphics*[width=0.49\textwidth,angle=0,clip]{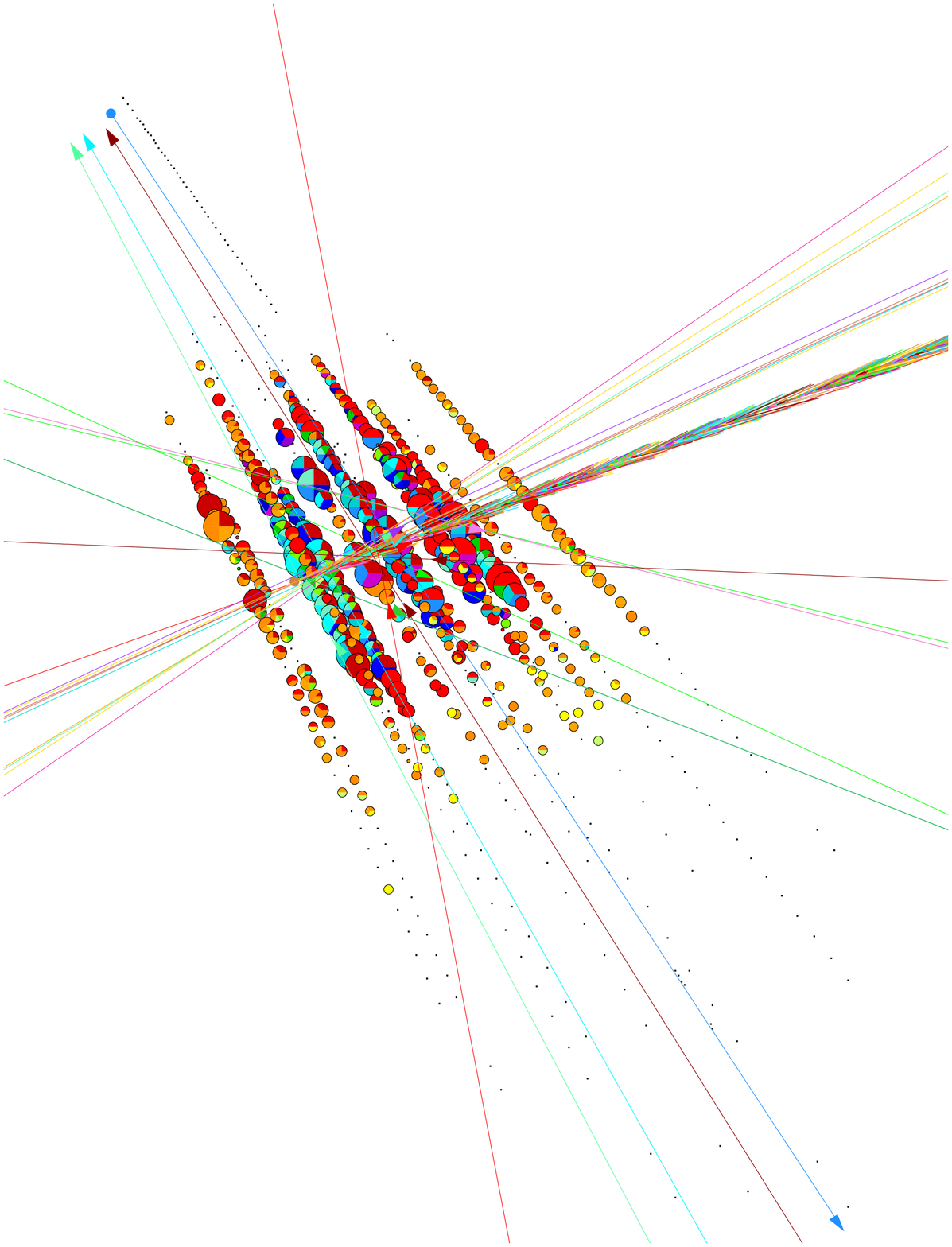}
\caption{\label {fig11}
(Left) one of the 1112 atmospheric neutrino event detected in the
  experimental array. (Right) a high-energy MC simulated event.
The size of the circles represent energy deposited in the
detector, the colors represent the time profile of photons
propagating in the ice which have scattered to the PMTs.}
\end{center}
\end{figure}

\section{Search for UHE Neutrinos}
The search for UHE neutrinos differs from the point source mainly in
two aspects.
First, the search for point sources is restricted to upward going events
of energy within the TeV scale, the UHE analysis extends the search
beyond PeV energies. Second, while point sources are localized in a
small region of the sky, UHE diffuse neutrinos are expected
to come from the entire visible sky (2$\pi$ sr), with a higher
probability to originate from the horizon.
Fig.~\ref{fig11} shows one of the 1112 upward going neutrino events
and an almost horizontal UHE simulated neutrino event with energy 
E $\sim 3 \times 10^{15}$ eV. Simulated events with energy above one
PeV deposit a substantial amount of light which is recorded by almost
all PMTs. In order to better separate background events from
signal events, we developed new variables which exploit the
information of the full waveforms.
\begin{figure}[h]
\begin{center}
\includegraphics*[width=0.49\textwidth,angle=0,clip]{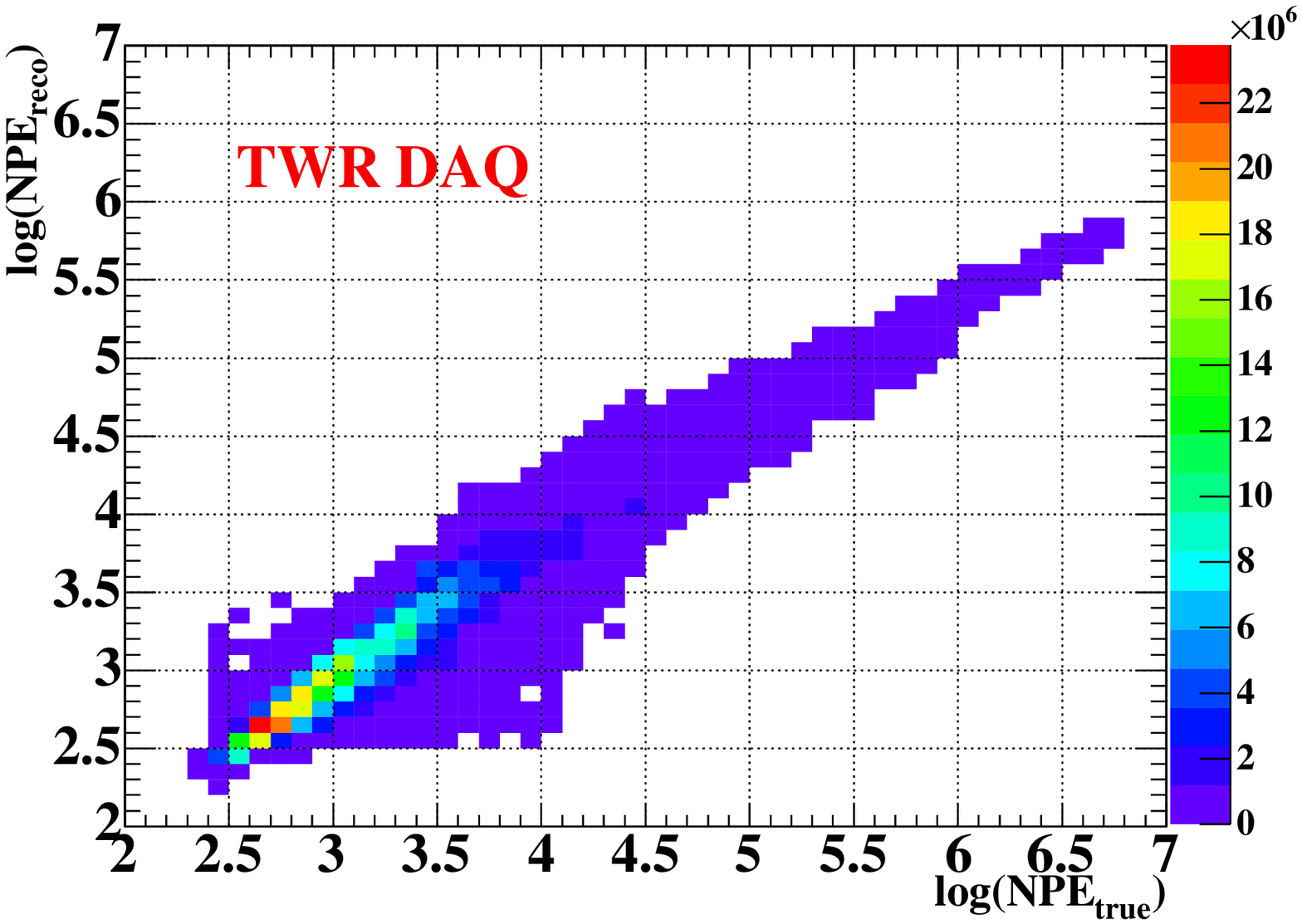}
\includegraphics*[width=0.49\textwidth,angle=0,clip]{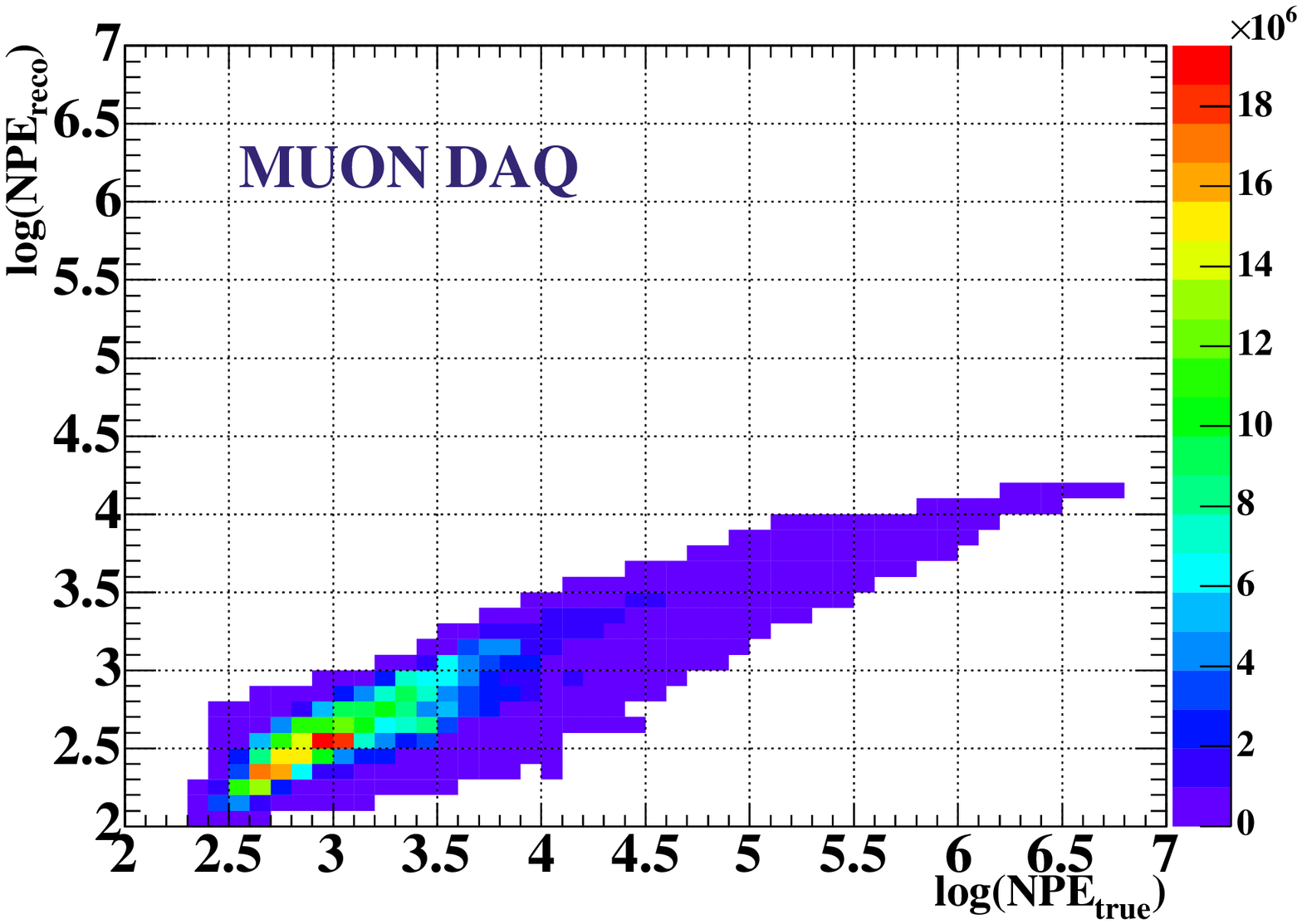}
\caption{\label {fig13}
Contour plots displaying the capability of correctly reconstructing
the total number of photon-electron injected into the detector (left)
for the {\tt TWR-DAQ} and (right) for the {$\mu$-{\tt DAQ}} system.}
\end{center}
\end{figure}
The capability of improved photo-electron reconstruction of
the {\tt TWR-DAQ} system can be seen in Fig.~\ref{fig13}, which displays a
contour plot in log$_{10}$-scale of the sum of the reconstructed number
of PE, NPE$_{reco}$ versus the NPE$_{true}$ generated by the signal MC
simulation.
While the {$\mu$-{\tt DAQ}} system quickly saturates on the order of
few thousands of PE's, the {\tt TWR-DAQ}, by using the
after-pulse information, can see the millions of PE's which are
typical for UHE events triggered by single muons.
Fig.~\ref{fig15} summarizes the results of several AMANDA searches for
diffuse neutrinos at different energy ranges using
{$\mu$-{\tt DAQ}} data.
The experimental limits assume a 1:1:1 ratio of neutrino flavors at the
Earth due to oscillation.
The dotted and dashed lines represent a sample of model
predictions{\small{\cite{Stecker92,Protheroe:1996,yos98,Sigl98,Stecker05}}},
adjusted for oscillation if necessary.
The horizontal solid lines represent limits on the integrated neutrino
flux from the diffuse analyses
with AMANDA-B10{\small\cite{diffuse-b10}}, with AMANDA-II{\small\cite{diffuse06}}, from the
cascade analysis{\small\cite{cascade}}, and from the UHE
analysis{\small\cite{UHEpaper,lisa06}}, with the NT200 neutrino
telescope{\small\cite{baikal}}, and at the highest energies from
ANITA-lite{\small\cite{barwick06}}and RICE{\small\cite{rice06}}. These are the most stringent flux
limits at 90\% C.L. for an E$^{-2}$ spectrum to date.
The AMANDA limits have been determined by analysing data collected with
the {$\mu$-{\tt DAQ}} system. However, as it has been shown above,
by performing a new analysis approach with {\tt TWR-DAQ} data, it is
possible to develop new techniques and to improve the current
experimental limits for the UHE neutrino search.
\begin{figure}[h]
\begin{center}
\includegraphics*[width=1.0\textwidth,angle=0,clip]{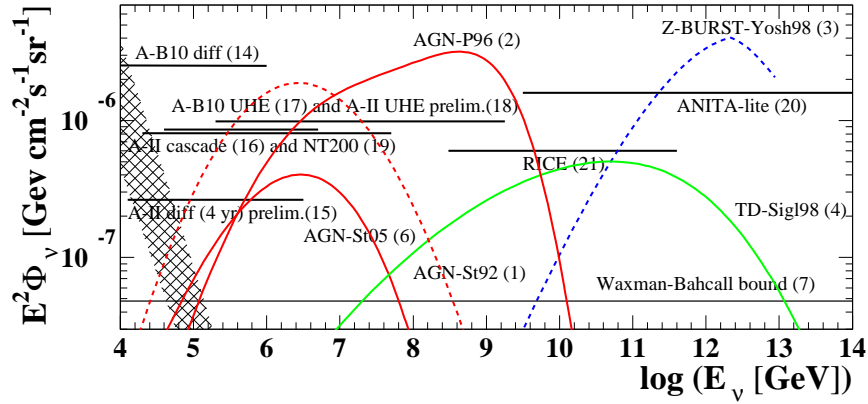}
\caption{\label {fig15}
Experimental limits on the integrated neutrino flux and detector
sensitivity for a neutrino flux of
E$^{-2}$ spectrum. The curves represents predictions from theoretical
models, the solid lines show current AMANDA (all-flavor limits),
NT200, ANITA and RICE experimental limits.}
\end{center}
\end{figure}

\section{Conclusion}
The point source search provides the first detailed
evaluation of the performance of the {\tt TWR-DAQ} system.
This analysis demonstrated that the 
{\tt TWR-DAQ} produces similar event rates and angular distribution as the data 
collected by the standard {$\mu$-{\tt DAQ}} system.
No statistically significant excess has been observed after performing a
full search of the northern hemisphere of the sky for localized event
clusters, therefore a flux limit based on {\tt TWR-DAQ} data is
calculated.
The AMANDA {\tt TWR} readout is now being incorporated into the
IceCube data acquisition system.
By exploiting the information of the full waveforms from the 
{\tt TWR-DAQ} system, it is possible to develop new analysis
techniques for an improved search for diffuse UHE neutrinos. Currently
the AMANDA experiment has placed the most stringent neutrino flux
limits to date, which can be further improved by analyses performed
with {\tt TWR-DAQ} data.

\section*{Acknowledgments}
The author acknowledges support from the U.S. National Science
Foundation (NSF) Physics Division, the NSF-supported TeraGrid systems at
the San Diego Supercomputer Center (SDSC) and the National Center for
Supercomputing Applications (NCSA), the Phi Beta Kappa Alumni in Southern
California for providing travel grants, the Astrophysics
Associates, Inc., Italian Ministry of Education, European Physical
Society, Ettore Majorana Foundation, and the Electron Tubes Ltd. for
providing the full scholarship at Erice.

\bibliographystyle{ws-procs9x6}

\end{document}